\documentstyle[12pt]{article}

\newcommand{\Lac}{\displaystyle{\biggl\{}}
\newcommand{\Rac}{\displaystyle{\biggr\}}}

\def\d{\delta}

\def\f{\phi}
\def\vf{\varphi}
\def\g{\gamma}

\def\j{\psi}

\def\l{\lambda}
\def\m{\mu}

\def\r{\rho}

\def\x{\xi}
\def\z{\zeta}
\def\D{\Delta}
\def\F{\Phi}
\def\G{\Gamma}

\def\Ome{\Omega}

\def\S{\Sigma}

\def\cf{{\cal F}}

\def\co{{\cal O}}

\def\cs{{\cal S}}

\def\inbar{\vrule height1.5ex width.4pt depth0pt}
\def\rlx{\relax\leavevmode}
\def\I{\leavevmode\hbox{\small1\kern-3.8pt\normalsize1}}
\def\openone{\leavevmode\hbox{\small1\kern-3.3pt\normalsize1}}
\def\Ione{\rlx{\rm 1\kern-2.7pt l}}
\def\Ik{\rlx{\rm I\kern-.18em k}}  
\def\IC{\rlx\leavevmode
             \ifmmode\mathchoice
                    {\hbox{\kern.33em\inbar\kern-.3em{\rm C}}}
                    {\hbox{\kern.33em\inbar\kern-.3em{\rm C}}}
                    {\hbox{\kern.28em\sinbar\kern-.25em{\rm C}}}
                    {\hbox{\kern.25em\ssinbar\kern-.22em{\rm C}}}
             \else{\hbox{\kern.3em\inbar\kern-.3em{\rm C}}}\fi}
\def\IP{\rlx{\rm I\kern-.18em P}}
\def\IR{\rlx{\rm I\kern-.18em R}}
\def\IN{\rlx{\rm I\kern-.20em N}}

\def\llsymbol#1{\@llsymbol{\@nameuse{c@#1}}}
\def\@llsymbol#1{\ifcase#1\or {}\or {'}\or {''}\or {'''}\or
   {''''}\or {'''''}\or  \else\@ctrerr\fi\relax}

\newcounter{contador}
\newcommand{\letra}{
   \stepcounter{equation}
   \setcounter{contador}{\value{equation}}
   \setcounter{equation}{0}
   \renewcommand{\theequation}{\thecontador.\alph{equation}}}
\newcommand{\antiletra}{
   \renewcommand{\theequation}{\arabic{equation}}
   \setcounter{equation}{\value{contador}}}

\newcommand{\ol}\overline
\newcommand{\ti}\tilde
\newcommand{\wt}\widetilde
\newcommand{\wh}\widehat
\newcommand{\bv}\breve
\newcommand{\dg}\dagger
\newcommand{\pari}{\stackrel{{P}}\longrightarrow}

\newcommand{\Ddd}{$D$$=$$1$$+$$2$}

\newcommand{\aand}{\;\;\;\mbox{and}\;\;\;}

\newcommand{\be}{\begin{equation}}
\newcommand{\ee}{\end{equation}}
\newcommand{\bl}{\begin{eqnarray}&}
\newcommand{\el}{&\end{eqnarray}}
\newcommand{\bq}{\begin{eqnarray}}
\newcommand{\eq}{\end{eqnarray}}

\newcommand{\sx}{\sigma_x}
\newcommand{\sy}{\sigma_y}
\newcommand{\sz}{\sigma_z}

\newcommand{\ov}{\overline}
\newcommand{\pa}{\partial}

\def\sl#1{\rlap{\hbox{$\mskip 1 mu /$}}#1}	
\def\Sl#1{\rlap{\hbox{$\mskip 3 mu /$}}#1}	
\def\SL#1{\rlap{\hbox{$\mskip 4.5 mu /$}}#1}	

\begin{document}

\title{\Large \bf Algebraic Renormalization of Parity-Preserving 
QED$_{3}$ Coupled to Scalar Matter I: Unbroken Case }

\author{{\it O. M. Del Cima}{\thanks{E-mail: 
oswaldo@cbpfsu1.cat.cbpf.br .}}~,~
{\it D. H. T. Franco}{\thanks{E-mail:
franco@cbpfsu1.cat.cbpf.br .}}~,~ {\it J. A. 
Helay\"el-Neto}~\\ and~{\it O. Piguet}{\thanks{On leave of absence from 
{\it D\'epartement de Physique Th\'eorique , Universit\'e de Gen\'eve,
24 quai E. Ansermet, CH--1211 Gen\`eve 4, Switzerland}. E-mail:
piguet@sc2a.unige.ch .}} {\thanks{Supported in part by the {\it Swiss 
National Science Foundation}.}}\\$\,$\\
{\normalsize Centro Brasileiro de Pesquisas F\'\i sicas (CBPF)} \\
{\normalsize Departamento de Teoria de Campos e Part\'\i culas (DCP)}\\
{\normalsize Rua Dr. Xavier Sigaud, 150 - Urca} \\
{\normalsize 22290-180 - Rio de Janeiro - RJ - Brazil.}\\$\,$\\
{\normalsize UGVA--DPT--1996--09--950}}

\date{}

\maketitle

\begin{abstract}
In this letter the algebraic renormalization method, which is 
independent of any kind of regularization scheme, is presented 
for the parity-preserving QED$_{3}$ coupled to scalar matter in 
the symmetric regime, where the scalar assumes vanishing vacuum 
expectation value, ${\langle}\vf{\rangle}$$=$$0$. The model 
shows to be stable under radiative corrections and anomaly free.
\end{abstract}

The study of gauge field theories in 3 space-time 
dimensions~{\cite{djt}} has been well-supported by a possible 
field-theoretical approach to describe some Condensed Matter phenomena, 
such as High-$T_{c}$ Superconductivity and Quantum
Hall Effect~{\cite{hightc,hall}}. Some Abelian models have been 
proposed in this direction, namely, the QED$_{3}$ and 
${\tau}_{3}$QED$_{3}$~{\cite{domavro,qedtau3}}. 

One of the interesting properties of 3-dimensional gauge 
field theories is the Landau gauge finiteness of non-Abelian 
Chern-Simons theories~{\cite{finiteness}}. 

The confinement of massive electrons in 3 space-time dimensions 
is a remarkable characteristic of this lower dimensional 
space~{\cite{maris}}. Recently, it was shown by using the Bethe-Salpeter 
equations that in a parity-preserving 
QED$_{3}$ there are bound states in electron-positron systems, 
positronium states~{\cite{bethesalp3}}.   

In a recent work~{\cite{e-pair}}, a parity-preserving 
QED$_{3}$ with spontaneous
breaking of a local $U(1)$-symmetry was proposed. The
breakingdown is accomplished by a sixth-power 
potential. It was shown that 
electrons scattered in {\Ddd}
can experience a mutual attractive interaction, 
depending on their spin states, where the intermediate bosons 
involved in such processes are a 
massive vector meson and a
Higgs scalar.
This attractive scattering potential comes from 
processes in which the
electrons are correlated in momentum space with 
opposite spin polarizations
($s$-wave state).   

One has still to study the renormalizability of this model, with the
$U(1)$ gauge invariance spontaneously broken as explained above.
However, the present letter is dedicated to the preliminary task of doing
that for the simpler case of unbroken gauge invariance. In this 
symmetric phase, the gauge boson remains massless. The same 
should occur for the fermion, since, in the broken phase, its mass is
completely generated by the Higgs mechanism.

But since a massless spinor might cause infrared singularities 
due to the presence of
super-renormalizable vertices involving the massless fields, we will add a
fermion mass term in order to avoid this problem -- which anyhow will not
appear in the physically interesting broken phase, where the fermion is
anyhow massive.

After a very brief summary of the model, we will show that its
parametrization is stable under small perturbations. This, together
with the proof of the absence of anomalies given in the final part of
the paper, will mean the
multipicative renormalizability of the theory.

The study of the renormalizability of the broken phase will be presented
in a forthcoming paper~\cite{broken-phase}.
\vspace{5mm}

The gauge invariant action for the parity-preserving QED$_{3}$\footnote{The 
metric adopted
throughout this work is $\eta_{mn}=(+,-,-)$; $m$,
$n$=(0,1,2). Note that slashed objects mean contraction 
with $\g$-matrices. The
latter are taken as $\g^m$$=$$(\sx,i\sy,-i\sz)$.} coupled to scalar 
matter~{\cite{e-pair}} in the $U(1)$-symmetric regime, 
${\langle}\vf{\rangle}$$=$$0$, is given by:
\bq
\S_{inv}\!\!\!\!&=&\!\!\!\!\int{d^3 x}      
\left\{ -{1\over4}
F^{mn}F_{mn}
+ i {\ov\j _+} {\SL{D}} {\j}_+ + i
{\ov\j _-} {\SL{D}} {\j}_- - m_0(\ov\j_+\j_+ - 
\ov\j_-\j_-) \;+\right.   
\nonumber\\
&&\left.
 -\;y (\ov\j_+\j_+ - 
\ov\j_-\j_-)\vf^*\vf + D^m\vf^* D_m\vf - \m^2\vf^*\vf - 
{\z\over2}(\vf^*\vf)^2 -
{\l\over3}(\vf^*\vf)^3 \right\} 
\;\;\;,
\label{inv}
\eq
where the mass dimensions of the parameters $m_0$, $\m$, 
$\z$, $\l$ and $y$ are
respectively ${1}$ ,${1}$, ${1}$, ${0}$ and ${0}$. 
The form of the potential is 
chosen such as to ensure the symmetric regime, 
where ${\langle}\vf{\rangle}$$=$$0$. 
Imposing that it must be bounded from below and yield only sable vacua, 
we get 
the following conditions on the parameters:
\be
\l>0 \;\;,\;\;\; \z<0 \aand \m^2 > {3\over 16} 
{\z^2\over \l} \;\;\;.
\label{parcond}
\ee
The covariant derivatives are defined as follows:
\be
{\SL{D}}\j_{\pm}\equiv(\sl{\pa} + iqg \Sl{A})\j_{\pm} 
\aand
D_{m}\vf\equiv(\pa_{m} + iQ g A_{m})\vf \;\;\;, 
\label{covder}
\ee
where $g$ is a coupling constant with dimension of 
(mass)$^{1\over2}$, and $q$
and $Q$ are the $U(1)$-charges of the fermions and 
scalar, respectively. In the
action (\ref{inv}), $F_{mn}$ is the usual field
strength for $A_m$, $\j_+$ and $\j_-$ are two kinds 
of fermions (the $\pm$
subscripts refer to their spin sign~{\cite{binegar}}) 
and $\vf$ is a complex
scalar. It should be noticed that in the action (\ref{inv}) a 
parity-preserving 
mass term for $\j_+$ and $\j_-$ has been added to the original action of 
ref.~{\cite{e-pair}} in order to avoid potential IR divergences which may 
be caused by the super-renormalizable interactions.  

The complete action, $\S$, we are considering here, is given by:
\bq
\S=\S_{inv}+\S_{gf}+\S_{ext}\;\;\;,\label{total}
\eq
where $\S_{gf}$ is the gauge-fixing action and $\S_{ext}$ is 
the action for the external sources:
\bq
\S_{gf}\!\!\!\!&=&\!\!\!\!\int{d^3 x}      
\left\{B{\pa}^mA_m + {\x\over2}B^2 + {\ov c}\Box c \right\} 
\;\;\;\;\;\;\;,\label{gf}
\eq
\bq
\S_{ext}\!\!\!\!&=&\!\!\!\!\int{d^3 x}      
\left\{ \ov\Ome_+s\j_+ - \ov\Ome_-s\j_- -
s\ov\j_+\Ome_+ + s\ov\j_-\Ome_- + \r^*s\vf + s\vf^*\r \right\} \;\;\;\;\;\;\;.
\label{ext} 
\eq
The gauge condition, the ghost equation and 
the antighost equation~\cite{bps-antigh} for (\ref{total}) read 
\letra
\bq
{\d\S\over\d B}\!\!\!\!&=&\!\!\!\!{\pa}^mA_m + \x B \;\;\;,\label{gaugecond}\\
{\d\S \over\d \ov c}\!\!\!\!&=&\!\!\!\!\Box c \;\;\;,\label{ghostcond}\\
-i{\d\S \over\d c}\!\!\!\!&=&\!\!\!\! \D_{\rm class}\;\;\;,\qquad\mbox{with:}
  \label{antighostcond}\\
\quad \D_{\rm class}\!\!\!\!&=&\!\!\!\! 
i\Box{\ov c} + q\ov\Ome_+\j_+ - q\ov\Ome_-\j_- +
q\ov\j_+\Ome_+ -q \ov\j_-\Ome_- - Q\r^*\vf - Q\vf^*\r \;\;\;.
\nonumber
\eq
\antiletra
Note that the right-hand sides being linear in the quantum fields,
will not be submitted to renormalization.
The QED$_{3}$-action\footnote{ For more details about 
QED$_{3}$ and
${\tau}_{3}$QED$_{3}$ as well as their applications, 
and some peculiarities of
parity and time-reversal in {\Ddd}, see refs.~{\cite{djt,domavro,qedtau3}}.}
(\ref{total}) is invariant under the reflexion 
symmetry $P$, whose action on the fields and external sources is fixed 
as below:
\begin{equation}\begin{array}{lll}
x_m &\ \pari\ & x_m^P=(x_0,-x_1,x_2)\;\;\;, \\
\j_{\pm} &\ \pari\ & \j_{\pm}^P=-i\g^1\j_{\mp}\;\;,\qquad 
\ov\j_{\pm}~\ \pari~\ \ov\j_{\pm}^P=i\ov\j_{\mp}\g^1\;\;\;, \\
A_m &\ \pari\ & A_m^P=(A_0,-A_1,A_2)\;\;\;,\\  
{\f} &\ \pari\ & {\f}^P=\f\ ,\qquad \f= \vf,\,c,\,\bar c,\,B \;\;\;, \\
\Ome_{\pm} &\ \pari\ & \Ome_{\pm}^P=-i\g^1\Ome_{\mp} \;\;,\qquad 
\ov\Ome_{\pm}~\ \pari\ ~\ov\Ome_{\pm}^P=i\ov\Ome_{\mp}\g^1\;\;\;,\\
\r &\ \pari\ & \r^P=\r \;\;\;. 
\end{array}\label{xp}\end{equation}
The ultraviolet and infrared 
dimensions\footnote{ We have to use a subtraction 
scheme which takes care of the presence of both massive and 
massless fields, subtracting off the UV divergences without introducing
spurious IR singularities. Such a scheme is the one of Lowenstein and 
Zimmermann~{\cite{low,troisieme}}. The UV and IR dimensions
mentioned here are those which are involved in
this formalism. The terms in the action, as well as all counterterms, 
are constrained to have UV dimension $\le3$ and IR dimension
$\ge3$.}, $d$ and $r$ respectively, 
as well as the ghost numbers, 
$\F\Pi$, and the Grassmann parity, $GP$, of all fields 
and sources are collected in Table~\ref{table1}. 

\begin{table}[hbt]
\centering
\begin{tabular}{|c||c|c|c|c|c|c|c|c|}
\hline
    &$A_m$ &$\vf$ &$\j_{\pm}$ &$c$ &${\ov c}$ &$B$ &$\r$ &$\Ome_{\pm}$  \\
\hline\hline
$d$ &${1/2}$ &${1/2}$ &1 &0 &1 &${3/2}$ &${5/2}$ &2  \\
\hline
$r$ &${1/2}$ &${3/2}$ &${3/2}$ &0 &1 &${3/2}$ &${5/2}$ 
&2  \\
\hline
$\F\Pi$&0 &0 &0 &1 &$-1$ &0 &$-1$ &$-1$  \\
\hline
$GP$&0 &0 &1 &1 &1 &0 &1 &0  \\
\hline
\end{tabular}
\caption[t1]{UV and IR dimensions, $d$ and $r$, ghost numbers, $\F\Pi$, and 
Grassmann parity, $GP$.}
\label{table1}
\end{table}

The BRS transformations are defined by:
\bq
&&s\vf=iQc\vf\;\;,\;\;\; s\vf^*=-iQc\vf^*\;\;\;, \nonumber\\
&&s\j_{\pm}=iqc\j_{\pm} \;\;,\;\;\; s\ov\j_{\pm}=-iqc\ov\j_{\pm}\;\;\;,
\nonumber\\
&&sA_m=-{1\over g}{\pa}_m c \;\;,\;\;\; sc=0 \;\;\;,\nonumber\\
&&s{\ov c}={1\over g}B \;\;,\;\;\; sB=0 \;\;\;,\label{BRS}
\eq
where $c$ is the ghost, ${\ov c}$ is the antighost and $B$ is the Lagrange 
multiplier field.

The BRS invariance of the action is expressed in a functional way by the 
Slavnov-Taylor identity
\bq
\cs(\S)=0 \;\;\;,\label{slavnovident}
\eq
where the Slavnov-Taylor operator  $\cs$ is defined,
acting on an arbitrary functional $\cf$, by
\bq
\cs(\cf)\!\!\!\!&=&\!\!\!\!\int{d^3 x}      
\left\{-{1\over g}{\pa}^m c {\d\cf\over\d A^m} + 
{1\over g}B {\d\cf\over\d {\ov c}} +
{\d\cf\over\d \ov\Ome_+}{\d\cf\over\d \j_+} -
{\d\cf\over\d \ov\Ome_-}{\d\cf\over\d \j_-} -
{\d\cf\over\d \Ome_+}{\d\cf\over\d \ov\j_+} +
{\d\cf\over\d \Ome_-}{\d\cf\over\d \ov\j_-} \;+\right.   
\nonumber\\
&&\left.
+\;{\d\cf\over\d\r^* }{\d\cf\over\d\vf } - 
{\d\cf\over\d\r}{\d\cf\over\d\vf^*}
\right\} \;\;\;\;\;\;\;.\label{slavnov} 
\eq
The corresponding linearized Slavnov-Taylor operator reads
\bq
\cs_\cf\!\!\!\!&=&\!\!\!\!\int{d^3 x}      
\left\{-{1\over g}{\pa}^m c {\d\over\d A^m} + 
{1\over g}B {\d\over\d {\ov c}} +
{\d\cf\over\d \ov\Ome_+}{\d\over\d \j_+} -
{\d\cf\over\d \ov\Ome_-}{\d\over\d \j_-} +
{\d\cf\over\d \j_+}{\d\over\d \ov\Ome_+} -
{\d\cf\over\d \j_-}{\d\over\d \ov\Ome_-} \;+\right.   
\nonumber\\
&&\left. 
-\;{\d\cf\over\d \Ome_+}{\d\over\d \ov\j_+} +
{\d\cf\over\d \Ome_-}{\d\over\d \ov\j_-} -
{\d\cf\over\d \ov\j_+}{\d\over\d \Ome_+} +
{\d\cf\over\d \ov\j_-}{\d\over\d \Ome_-} +
{\d\cf\over\d\r^* }{\d\over\d\vf } +
{\d\cf\over\d\vf }{\d\over\d\r^* } \;+\right.   
\nonumber\\
&&\left.
-\;{\d\cf\over\d\r }{\d\over\d\vf^* } -
{\d\cf\over\d\vf^* }{\d\over\d\r } 
\right\} \;\;\;\;\;\;\;.\label{slavnovlin} 
\eq
The following nilpotency identities hold:
\letra
\bq
\cs_\cf\cs(\cf)=0 \;\;,\;\;\;\forall\;\cf \;\;\;,\label{nilpot1} \\
\cs_\cf\cs_\cf=0 \;\;\;{\mbox{if}} \;\;\;\cs(\cf)=0\;\;\;. \label{nilpot3}
\eq
\antiletra
In particular:
\bq
\left(\cs_\S\right)^2 = 0\;\;\;,
\label{nilpot2}\eq
since the action $\S$ obeys the Slavnov-Taylor identity.
The operation of $\cs_{\S}$ over the fields and the external sources is 
given by
\bq
&&\cs_{\S}\f=s\f \;\;,\;\;\;\f=\j_{\pm},\;\ov\j_{\pm},\;\vf,\;\vf^*,
\;A_m,\;c,\;
{\ov c}\!\!\aand\!\! B \;\;\;,\nonumber\\ 
&&\cs_{\S}\ov\Ome_+={\d\S\over\d \j_+} \;\;,\;\;\;
\cs_{\S}\ov\Ome_-=-{\d\S\over\d \j_-} 
\;\;\;,\nonumber\\
&&\cs_{\S}\Ome_+=-{\d\S\over\d \ov\j_+}  \;\;,\;\;\; 
\cs_{\S}\Ome_-={\d\S\over\d \ov\j_-} \;\;\;,\nonumber\\
&&\cs_{\S}\r^*={\d\S\over\d\vf }  \;\;,\;\;\;
\cs_{\S}\r=-{\d\S\over\d\vf^* }\;\;\;.\label{operation1}
\eq
In order to study the stability~{\cite{brs}} of 
the action (\ref{total}) under the 
radiative corrections, one has to find the most general counterterm, 
$\S^c$, 
satisfying the following condition of BRS invariance:
\bq
\cs_{\S}\S^c=0 \;\;\;\;. \label{stabcond}
\eq
$\S^c$ is an integrated local polynomial in the fields and its 
derivatives with UV dimension $\le3$, IR dimension $\ge3$  
and with vanishing ghost number.
 It has to be invariant under the $P$-symmetry given by 
Eqs.(\ref{xp}), and it has also to satisfy the conditions  
\bq
{\d\S^c\over{\d B} }=0 \;\;,\;\;\;{\d\S^c\over{\d{\ol c}}}=0 
\;\;,\;\;\; {\d\S^c\over{\d c}}=0 \;\;\;\;, \label{suplcond}
\eq
which follow from the conditions (\ref{gaugecond} -- 
\ref{antighostcond}), and, moreover:
\bq
W_{\rm rigid} \S^c=0\;\;\;\;, \label{crigidcond}
\eq
where $W_{\rm rigid}$ is the Ward operator of rigid  symmetry defined by
\bq
W_{\rm rigid}\!\!\!\!&=&\!\!\!\!\int{d^3 x}      
\left\{
q\j_+{\d\over\d \j_+} +
q\j_-{\d\over\d \j_-} -
q\ov\j_+{\d\over\d \ov\j_+} -
q\ov\j_-{\d\over\d \ov\j_-} +
Q\vf{\d\over\d\vf} -
Q\vf^*{\d\over\d\vf^*} \;+\right.   
\nonumber\\
&&\left.
+\;q\Ome_+{\d\over\d \Ome_+} +
q\Ome_-{\d\over\d \Ome_-} -
q\ov\Ome_+{\d\over\d \ov\Ome_+} -
q\ov\Ome_-{\d\over\d \ov\Ome_-} +
Q\r{\d\over\d\r} -
Q\r^*{\d\over\d\r^*}
\right\}\;\;\;. \label{wrigid} 
\eq
Eq.(\ref{crigidcond}) follows 
from the rigid $U(1)$ invariance of the 
action\footnote{ Rigid invariance itself 
follows from the antighost equation
(\ref{antighostcond}) and from the validity of the Slavnov-Taylor 
identity (\ref{slavnovident}).}:
\bq
W_{\rm rigid} \S=0 \;\;\;\;. \label{rigidcond}
\eq
We find that the most general invariant counterterm
$\S^c$, i.e. the most
general field polynomial of UV and IR dimensions 
bounded by $d$$\leq$$3$ and $r$$\geq$$3$, with 
ghost number zero, respecting $P$-symmetry, (\ref{xp}) and the 
conditions displayed in Eqs.(\ref{stabcond}), 
(\ref{suplcond}) and (\ref{crigidcond}), is given by
an arbitrary superposition of the following expressions:
\be\begin{array}{l}
\Lac  F^{mn}F_{mn}\ ,\ \ 
i({\ov\j _+} {\SL{D}} {\j}_+ + {\ov\j _-} {\SL{D}} {\j}_- )\ ,\ \  
(\ov\j_+\j_+ - \ov\j_-\j_-) \ ,\ \ \\ \phantom{\Lac}   
(\ov\j_+\j_+ - \ov\j_-\j_-)\vf^*\vf \ ,\ \ D^m\vf^* D_m\vf \ ,\ \ 
\vf^*\vf  \ ,\ \ (\vf^*\vf)^2\ ,\ \  (\vf^*\vf)^3 \Rac \;\;\;. 
\end{array}\label{finalcount}\ee
The BRS consistency condition in the sector of ghost number zero, given by 
Eq.(\ref{stabcond}), constitutes a cohomology problem due to the nilpotency 
(\ref{nilpot2}) of the linearized Slavnov-Taylor operator (\ref{slavnovlin}). 
Its solution can always be written as a sum of a trivial cocycle 
$\cs_{\S}\wh\S$, where $\wh\S$ has ghost number $-1$, and a nontrivial
part $\S_{\rm phys}$ 
belonging to the cohomology of $\cs_{\S}$ (\ref{slavnovlin}) 
in the sector of ghost number zero, i.e. which cannot be written 
as a $\cs_{\S}$-variation:
\be
\S^c=\S_{\rm phys} + \cs_{\S}\wh\S \;\;\;. \label{split}
\ee
One checks indeed that the general invariant counterterm, expanded 
in the basis (\ref{finalcount}), admits the representation 
(\ref{split}), with
\letra
\bq
\S_{\rm phys} \!\!\!\!&=&\!\!\!\! z_g\left(g{\pa\over{\pa g}} - N_A + N_B -
2\x{\pa\over{\pa \x}}\right)\S + z_{m_0}\;m_0{\pa\S\over{\pa m_0}}\;+ 
\nonumber\\
&&+\; z_y\;y{\pa\S\over{\pa y}} + z_{\m^2}\;\m^2{\pa\S\over{\pa \m^2}} + 
 z_{\z}\;\z{\pa\S\over{\pa \z}} + z_{\l}\;\l{\pa\S\over{\pa \l}}  \;\;\;,\\
\cs_{\S}\wh\S \!\!\!\!&=&\!\!\!\!  \cs_{\S} \int{d^3 x} \left[ 
  z_{\j}\left(\ov\j_+\Ome_+ - \ov\Ome_+\j_+ -  \ov\j_-\Ome_- 
  + \ov\Ome_-\j_- \right) 
 + z_{\vf} \left( \r^*\vf - \vf^*\r \right)    \right] \nonumber\\
\!\!\!\!&=&\!\!\!\! z_{\j}\left( N_{\j_+} + N_{\ov\j_+} +N_{\j_-} 
  + N_{\ov\j_-} - N_{\Ome_+} - N_{\ov\Ome_+} - N_{\Ome_-} 
  - N_{\ov\Ome_-} \right)\S  \nonumber\\
&&+\;  z_{\vf}\left( N_{\vf} + N_{\vf^*} - N_{\r} - N_{\r^*} \right)\S 
  \;\;\;, \label{countcount}
\eq
\antiletra
where the counting operators are defined by
\be
N_{\f}=\int{d^3 x} \;\f\;{\d\over\d \f} \;\;,\;\;\;\f=\j_{\pm},\;\ov\j_{\pm}
,\;\Ome_{\pm},\;\ov\Ome_{\pm},\;\vf,\;\vf^*,\;\r,\;\r^*,\;A_m\!\!
\aand\!\! B \;\;\;. \label{countdef}
\ee 
This way of writing the counterterm makes explicit the separation
between the physical counterterms, on the one hand, which 
amount to the  renormalization of the physical masses and coupling 
constants $m_0$, $\m$, $g$, $y$, $\zeta$, $\lambda$, and the trivial 
ones, on the other hand, which correspond
 to the unphysical renormalization of the amplitudes of the
fields $\psi_\pm$ and $\varphi$ -- the other field 
renormalizations not being independent. The form of the classical 
action $\S$ (\ref{total}), taken as a $P$-invariant solution of 
the functional identities  expresssing the  various symmetries of 
the theory, is thus stable under small perturbations, the general 
solution in a neighbourhood of $\S$ being  obtained through an
arbitrary  variation of the parametrization.
\vspace{5mm}

At the quantum level the vertex functional $\G$, 
which coincides with the classical action (\ref{total}) 
at order 0 in $\hbar$:
\be
\G=\S + {\co}(\hbar) \;\;\;,\label{vertex}
\ee
has to satisfy the constraints
\letra
\bq
&&{\d\G\over\d B}={\pa}^mA_m + \x B \;\;\;,\label{qgaugecond}\\
&&{\d\G\over\d\ol c}={\Box c} \;\;\;,\label{qghost}\\
&&-i{\d\G\over\d c} = \D_{\rm class}\;\;\;,\label{qantighost}\\
&&W_{\rm rigid} \G=0 \;\;\;, \label{qrsuplcond}
\eq
\antiletra
where $W_{\rm rigid}$ has already been defined by equation 
(\ref{wrigid}) and Eqs.(\ref{qgaugecond} -- \ref{qantighost})
are the quantum extension of Eqs.(\ref{gaugecond} -- \ref{antighostcond}).

According to the Quantum Action Principle~{\cite{sopi,qap}} the 
Slavnov-Taylor identity (\ref{slavnovident}) gets a quantum breaking
\be
\cs(\G)=\D \cdot \G = \D + {\co}(\hbar \D)\;\;\;, \label{slavnovbreak}
\ee
where $\D$ is an integrated local functional with ghost number 1 and
dimension 3.

The nilpotency identity ({\ref{nilpot1}) together with
\be
\cs_{\G}=\cs_{\S} + {\co}(\hbar)
\ee
implies the following consistency condition for the breaking $\D$:
\be
\cs_{\S}\D=0 \;\;\;.\label{breakcond1}
\ee
Other constraints on $\D$ follow from the constraints 
(\ref{qgaugecond} -- \ref{qrsuplcond}) and from the algebra
\letra
\bq
&&{\d\cs(\cf)\over\d B} - \cs_{\cf}\left({\d\cf\over\d B}- {\pa}^mA_m - \x 
B \right)={1\over g}\left({\d\cf\over\d\ol c } - {\Box c}\right)\;\;\;,
\label{fcond1}\\
&&{\d\cs(\cf)\over\d\ol c} + \cs_{\cf}{\d\cf\over\d\ol c }=0\;\;\;,
\label{fcond2} \\
&&-i\int d^3x \frac{\d}{\d c} \cs(\cf) + 
  \cs_{\cf}\int d^3x \left(-i\frac{\d}{\d c}\cf - \D_{\rm class}\right)
   =W_{\rm rigid} \cf \;\;\;, \label{fcond3}\\
&&W_{\rm rigid}\cs(\cf) - \cs_{\cf}W_{\rm rigid} \cf =0 \;\;\;,
  \label{fcond4}\\[3mm]
&&\mbox{($\cf$ arbitrary functional of ghost number zero)} \;\;\;.\nonumber
\eq
\antiletra
These constraints on the breaking $\D$ read:
\letra
\bq
&&{\d\D\over\d B}=0  \;\;\;,\label{breakcond2}\\
&&{\d\D\over\d\ol c}=0 \;\;\;,\label{breakcond3}\\
&&\int d^3x \frac{\d}{\d c} \D=0 \;\;\;,\label{breakcond4}\\[3mm]
&&W_{\rm rigid} \D=0 \;\;\;.\label{breakcond5}
\eq
\antiletra
The Wess-Zumino consistency condition (\ref{breakcond1}) constitutes a 
cohomology problem like in the zero ghost number case (\ref{stabcond}). 
Its solution can always be written as a sum of a trivial cocycle 
$\cs_{\S}{\wh\D}^{(0)}$, where ${\wh\D}^{(0)}$ has ghost number $0$, 
and of nontrivial elements belonging to the cohomology of $\cs_{\S}$ 
(\ref{slavnovlin}) in the sector of ghost number one:
\be
\D^{(1)} = {\wh\D}^{(1)} + \cs_{\S}{\wh\D}^{(0)} \;\;\;, 
\label{breaksplit}
\ee
where $\D^{(1)}$ must be even under $P$-symmetry and obey the conditions 
imposed by Eqs. (\ref{breakcond2} -- \ref{breakcond5}). The trivial cocycle
 $\cs_{\S}{\wh\D}^{(0)}$ can be absorbed into the vertex functional $\G$ 
as a noninvariant integrated local couterterm $-{\wh\D}^{(0)}$. 
On the other hand, a nonzero $\D^{(1)}$ would represent an anomaly.

Considering the condition (\ref{breakcond4}), to be satisfied by 
(\ref{breaksplit}), it can be concluded that  
\be
\D^{(1)} = \int{d^3 x} \;K^{(0)}_m\;{\pa}^mc \;\;\;. \label{anomaly}
\ee
By analyzing the Slavnov-Taylor operator $\cs_{\S}$ (\ref{slavnovlin}) and 
the Wess-Zumino consistency condition (\ref{breakcond1}), one sees that the 
breaking 
$\D^{(1)}$ has UV and IR dimensions bounded by  
$d$$\leq$${7\over2}$ and $r$$\geq$$2$. Therefore, the dimensions of
$K^{(0)}_m$ 
must be bounded by $d$$\leq$${5\over2}$ and $r$$\geq$$1$, it has
ghost number $0$, and due to Eq.(\ref{breakcond1}) and Eqs.
(\ref{breakcond2} -- \ref{breakcond3}), it must respect the 
conditions
\be
{\d K^{(0)}_m\over\d B}=0 \aand {\d K^{(0)}_m\over\d\ol c}=0
\;\;\;.\label{condk0}\\
\ee 
Now, rewriting $K^{(0)}_m$ as a linear combination 
\be
K^{(0)}_m = {\sum_{i=1}^{7} }\; a_i \; K^{(0)i}_m \;\;\;, \label{lincombk0}
\ee
where 
\bq
&&K^{(0)1}_m = A_m A^nA_n \;\;,\;\;
K^{(0)2}_m = A_m (A^nA_n)^2 \;\;,\;\;
K^{(0)3}_m = A_m (\ov\j_+\j_+ - \ov\j_-\j_-)\;\;, \nonumber\\
&&K^{(0)4}_m = A_m A^nA_n \vf^*\vf  \;\;,\;\; 
K^{(0)5}_m = A_m (\vf^*\vf)^2 \;\;,\;\;
K^{(0)6}_m = A_m \vf^*\vf \;\;, \nonumber\\
&&K^{(0)7}_m = \ov\j_+\g_m\j_+ + \ov\j_-\g_m\j_- \;\;\;,  
\eq
and solving all the conditions it has to fulfil, we 
can easily show, with the help of Eqs.(\ref{operation1}), 
that there exist local functionals ${\wh\D}^{(0)i}$ such that 
\be 
\int d^3x \;K^{(0)i}_m\;\pa^m c = \cs_{\S}{\wh\D}^{(0)i}\ ,\ \ 
     i=1,\cdots,7\;\;\:.
\label{lincomd0}
\ee
This means ${\wh\D}^{(1)}$$=$$0$ in (\ref{breaksplit}), which
implies the implementability of the Slavnov-Taylor identity to every
order through the absorbtion of the noninvariant counterterm 
$-{\sum_{i}} a_i{\wh\D}^{(0)i}$.

Of course, invariant counterterms may still be arbitrarily added at each
order. However the result of the discussion on the stability of 
the classical theory shows that these counterterms correspond to a
renormalization of the parameters of the theory. Their coefficients have
to be fixed by suitable normalization conditions.

In conclusion, we have shown the renormalizability and absence of gauge
anomaly for the parity-preserving QED${}_3$ coupled to scalar matter 
in the symmetric phase.

\small

\subsection*{Acknowledgements}
The authors express their gratitude to Prof. S.P. Sorella for 
patient and helpful discussions. One of 
the authors (O.M.D.C.) dedicates 
this work to his wife,
Zilda Cristina, and to his daughter, Vittoria, who was 
born in 20 March 1996. O.M.D.C. and D.H.T.F. thank to the 
{\it High Energy Section} of the {\it ICTP - Trieste - Italy}, 
where this work was done, for the kind hospitality and 
financial support, and to its Head, Prof. S. Randjbar-Daemi. O.P. 
thanks the CBPF, the Physics Department of the Federal University 
of Esp\'\i rito Santo (Brazil) and his collegues from both institutions 
for their very kind hospitality.
Thanks are also due to the Head of CFC-CBPF,
Prof. A.O. Caride, and Dr. R. Paunov, for encouragement. 
CNPq-Brazil is 
acknowledged for
invaluable
financial help.


\end{document}